\documentstyle[aps,pra,subfigure,epsfig,amsmath,amssymb]{revtex}
\begin{document}
\draft

\title{ Momentum distribution of an interacting Bose-condensed gas
at finite temperature}
\author{ Anna Minguzzi, Patrizia Vignolo and Mario P. Tosi}
\address{  Istituto Nazionale di Fisica della Materia and Classe di Scienze,
Scuola Normale Superiore,
Piazza dei Cavalieri 7, I-56126 Pisa, Italy}
\maketitle
\begin{abstract}
We use a semiclassical two-fluid model to study the momentum distribution
of a Bose-condensed gas with repulsive interactions inside a harmonic
trap at finite temperature, with specific focus on atomic hydrogen.
We give particular attention to the average kinetic energy, which is almost
entirely associated with the thermal cloud. A non-linear dependence
of the kinetic energy on temperature is displayed, affording a precise 
way to assess the temperature of the gas. We also show that the kinetic 
energy increases with the strength of the interactions, reflecting an 
enhanced rate of depletion of the condensate with increasing temperature.  
\end{abstract}

\section{Introduction}
The observation in momentum space of Bose-Einstein condensation in a 
trapped gas with repulsive interactions presents at least in
principle some advantage over observation in coordinate space.
Since the condensate belongs to a single quantum state,
a broadening of its density profile
implies a narrowing of its momentum distribution.
At the same time the width of the momentum distribution of the thermal 
cloud is in essence unchanged, the role of quantal indeterminacy
being negligible for this component.

The momentum distribution of a dilute Bose-condensed gas has been
measured both on trapped and expanding vapours of $^{23}$Na at $T\simeq0$
\cite{ketterle_S_1} and on trapped vapours of spin-polarized $^1$H 
atoms at finite temperature \cite{bec_h}. 
The former experiment has been successfully
interpreted by means of a Thomas-Fermi model \cite{baym,yvan_exp}.
In the present work we are concerned with the latter experiment,
in which the momentum distribution of the hydrogen cloud was inferred from
the Doppler-sensitive part of the 1$S$-2$S$ two-photon absorption
spectrum: the measured distribution 
shows a small condensate peak on top of a broad thermal
component. 
Cold collisions in the high-density condensate broaden
the peak associated with it~\cite{cold_collision_shift,killian}.
Here we are instead interested in extracting
thermodynamic information on the system from the thermal part
of the spectrum.

We use a semiclassical two-fluid model, which was earlier developed 
to account for the release energy in expanding clouds
of $^{87}$Rb as a function of temperature \cite{conti97}.
With this model we evaluate the momentum
distribution of an interacting Bose gas at equilibrium
under harmonic
confinement. 
The results for a choice of parameters corresponding to
the experiment on hydrogen agree with the measured
condensate fraction within
its error bars. 
We place particular attention on the average kinetic energy of the gas,
displaying its temperature dependence and the role of the
interactions.

Before proceeding we should mention that, in addition to the 
Doppler-sensitive part of present interest, the measured spectrum
also includes a Doppler-free contribution. The thermal 
portion of this Doppler-free spectrum has been interpreted
within an out-of-equilibrium picture invoking large
fluctuations in the particle density\cite{cote}. 
However, the
contribution of such fluctuations to the Doppler-sensitive thermal
spectrum is expected to be of the same order as for the Doppler-free
line, making it smaller than the Doppler
broadening by two orders of magnitude.

\section{The model and its justification}
In our description of the Bose 
 gas with repulsive interactions, 
confined in a harmonic trap at finite temperature
$T$, we adopt the semiclassical Hartree-Fock (HF) scheme
for the thermal cloud and the Thomas-Fermi (TF) approximation
for the condensate.

The choice of a mean field theory is justified by the fact that in
current experiments
the gas is in a very dilute regime, corresponding 
in the trap to the condition
$N^{1/6}a/a_{ho}\ll 1$ \cite{giorgini_scaling} where
$N$ is the number of particles, $a$ is the $s$-wave scattering
length and $a_{ho}=(\hbar/m\omega)^{1/2}$ is the harmonic oscillator length.
Here $\omega=(\omega_{\perp}^2\omega_z)^{1/3}$
is the geometrical average of the harmonic oscillator frequencies
characterizing the axially symmetric confining potential 
$V_{ext}({\bf r})=m(\omega_{\perp}^2r_{\perp}^2+\omega_z^2z^2)/2$. 
In particular, in the hydrogen experiment $N^{1/6}a/a_{ho}\simeq0.001$.

In the presence of confinement the HF scheme has been shown to
provide a simple and accurate description of 
the excitation spectrum \cite{muntsa}, yielding
results which are comparable with the predictions of
the full Bogoliubov theory.
The semiclassical approximation is valid for a bosonic
cloud when its temperature is appreciably
larger than the harmonic-oscillator spacing, 
$k_BT\gg\hbar\omega_{\perp,z}$. 
In particular, in the hydrogen experiment $k_BT\simeq 270\hbar\omega_{\perp}
\simeq10^5 \hbar\omega_z$.
For the specific case
of trapped atomic vapours, the semiclassical HF approximation 
has also been tested by a Monte Carlo calculation \cite{markus_hf},
performed with a choice of parameters corresponding to the $^{87}$Rb
experiments \cite{cornell_thermod}. 

The TF approximation for the condensate well describes
the gas with repulsive interactions
in the strong
coupling limit $N_c\, a/a_{ho}\gg 1$, {\it i.e.} when the
number $N_c$ of atoms in the condensate is large. 
This approximation is therefore valid everywhere except in the critical
region. In the hydrogen experiment one has $N_ca/a_{ho}\simeq1.6\cdot 10^4$.

On all these counts, therefore, the use of the semiclassical
two fluid model is fully justified in the calculations that
we present below, except in the immediate neighbourhood
of the critical temperature as remarked just above.
The momentum distributions of the thermal cloud $f_T({\bf p})$
and of the condensate $f_{c}({\bf p})$ are then given
by the Bose distribution in an effective potential and by the square 
modulus of the Fourier transform
of the wavefunction of the condensate:
\begin{equation}
f_T({\bf p})=\int \left\{\exp{\left[
\dfrac{1}{k_BT}\left(\dfrac{p^2}{2m}+V_{eff}
({\bf r})-\mu\right)\right]}-1\right\}^{-1}{\rm d}^3{\bf r}\;,\label{model} 
\end{equation}
and
\begin{equation}
f_{c}({\bf p})=\left|\int\phi_c({\bf r})e^{-i{\bf p}\cdot{\bf r}}
{\rm d}^3{\bf r}\right|^2\;.
\end{equation}
Here $\phi_c({\bf r})=\sqrt{n_c({\bf r})}$, the square root of the
density profile of the condensate. In Eq.~(\ref{model})
the effective potential acting on the thermal cloud is given by
\begin{equation}
V_{eff}({\bf r})=V_{ext}({\bf r})+2gn_c({\bf r})+2gn_T({\bf r})\;,
\end{equation}
$n_T({\bf r})$ being the density profile of the thermal cloud.
Finally, we have
\begin{equation}
n_c({\bf r})=\dfrac{1}{g}\,\left[\mu-V_{ext}({\bf r})-2gn_T({\bf r})\right]\,
\theta(\mu-V_{ext}({\bf r})-2gn_T({\bf r}))
\end{equation}
and
\begin{equation}
n_T({\bf r})=\int \left\{\exp{\left[
\dfrac{1}{k_BT}\left(\dfrac{p^2}{2m}+V_{eff}
({\bf r})-\mu\right)\right]}-1\right\}^{-1}
\dfrac{{\rm d}^3{\bf p}}{(2\pi)^3}.
\end{equation}
The coupling constant $g$ is fixed by the $s$-wave
scattering length $a$ through the relation 
$g=4\pi\hbar^2 a/m$, and the chemical potential 
$\mu$ is fixed
by the total number $N$ of atoms,
\begin{equation}
\int \left[n_c({\bf r})+n_T({\bf r})\right]{\rm d}^3{\bf r}=N\;.
\end{equation}

%In the dilute limit where $n_T\ll n_c$ the effective potential $V_{eff}$
%can be written in an analytic form
%\[V_{eff}=V_{ext}+2(\mu-V_{ext})\,\theta(\mu-V_{ext}).\]
%This double parabola model will be useful to understand qualitatively 
%the density profiles of the thermal cloud.

For temperatures above the critical condensation temperature 
the expression (\ref{model}) is equivalent to
that developed by Chou {\it et al.}~\cite{chou2}.

\section{Momentum distributions}
\label{sez2}
In the experiments of Fried et al. \cite{bec_h} on atomic hydrogen 
the momentum distribution has been inferred from
the Doppler-sensitive 
spectrum as measured along the weakly confining axis of the trap.
We accordingly  report in Figures 1 and 2
the distributions for the thermal cloud and for the condensate 
as functions of the axial momentum $p_z$:
\begin{equation}
F_{T,c}(p_z)=\int\frac{{\rm d}^2p_{\perp}}{(2\pi)^2}
f_{T,c}({\bf p}).
\end{equation}
The units of length and momentum are 
$a_{ho\perp}=(\hbar/m\omega_{\perp})^{1/2}$ and
$p_{ho\perp}=(\hbar m\omega_{\perp})^{1/2}$, with
the values of $m$ and $\omega_{\perp}$ of the experiment
\cite{bec_h}. We have taken $\omega_\perp=2\pi \cdot 3.9$ kHz and
$\omega_z=2\pi \cdot 10.2$ Hz.

\begin{figure}
\centerline{\epsfig{file=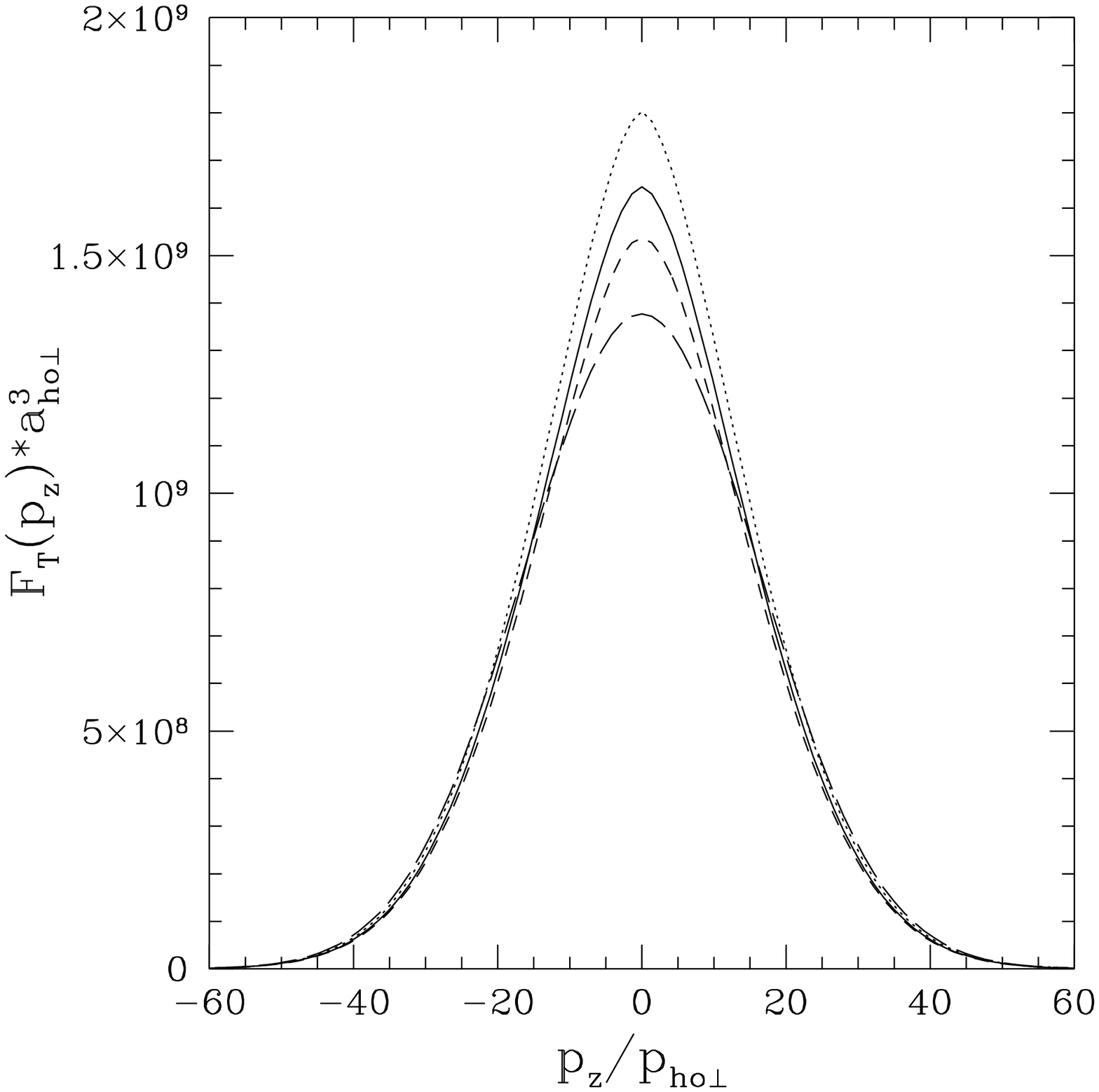,width=12cm}}
\caption{Axial momentum distribution of the thermal component of
a weakly interacting Bose gas with $N=10.5\cdot10^9$
at $T=50.3\,\mu$K. The scattering length is $a=20a_H$
(dotted curve) and $a=a_H$
(continuous curve) with $a_H=6.48\cdot 10^{-2}\,$nm. 
These two curves are compared with the
momentum distribution of the non-interacting Bose gas
(dashed curve) and of the classical gas (long-dashed curve). The
latter has the same number of particles as the non-interacting
quantal gas.}
\label{fig1}
\end{figure}

Figure \ref{fig1} illustrates both the role of statistics
and the effect of the interactions on  the axial
momentum distribution of the
thermal component.
The predictions of the semiclassical model for a weakly interacting
gas, with a scattering length chosen equal to that
of $^1$H ($a_H=6.48\cdot 10^{-2}$ nm from accurate calculations
by Jamieson {\it et al.} \cite{jamieson}) or arbitrarily
increased to $20\,a_H$ for illustrative purposes, are compared
with each other and with that for a non-interacting Bose gas.
The value of the temperature is $T=50.3\,\mu$K,
as will be discussed in Sect.~\ref{sez4} below.
Figure \ref{fig1} also reports the momentum distribution of
the classical gas at the same temperature.

It is evident from Figure \ref{fig1} that quantum statistics 
increases the population of the states 
at low momentum, as expected. The main effect
of the interactions is instead a depletion of the condensate,
leading to increases in the population and 
in the width of the momentum
distribution of the thermal cloud.

\begin{figure}
\centerline{\epsfig{file=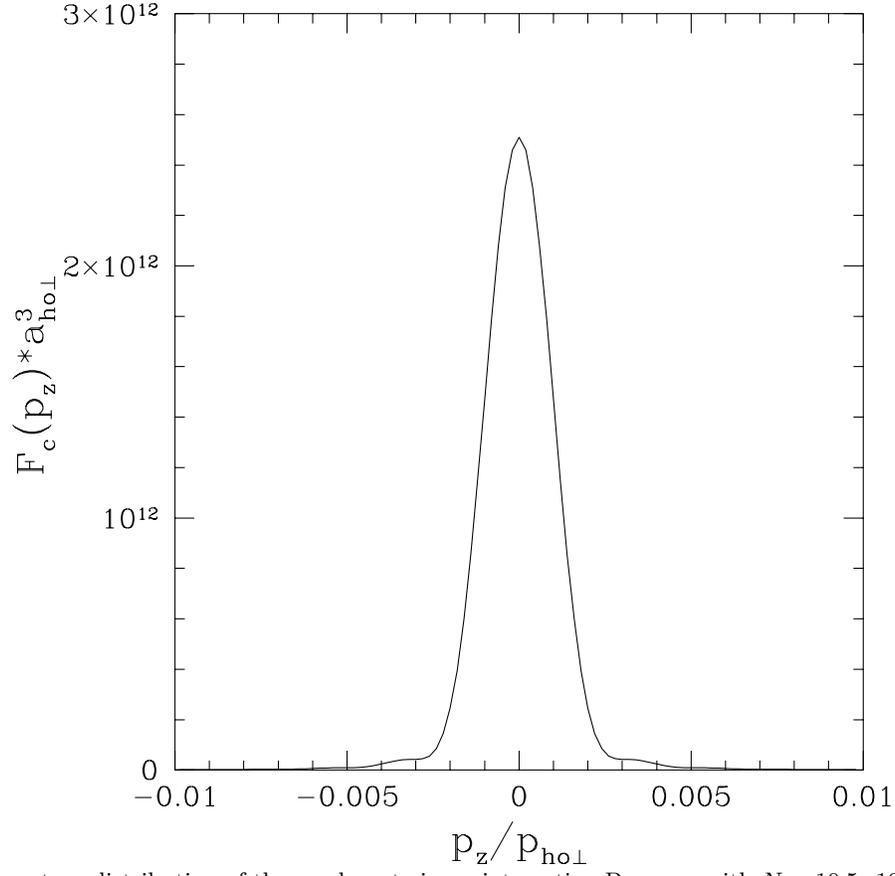,width=12cm}}
\caption{Axial momentum distribution of the condensate in
an interacting Bose gas with $N=10.5\cdot10^9$
at $T=50.3\mu$K. The scattering length is $a=a_H$.}
\label{fig2}
\end{figure}

The momentum distribution of the condensate at the same temperature
as coming from the
Thomas-Fermi model is shown in Fig.~\ref{fig2}.

\section{Kinetic energy} 
\label{sez4}
We proceed to study the momentum
distribution of 
an interacting Bose gas as a function of temperature 
by focusing on its second moment, that is on the kinetic energy of the
trapped cloud. This is the quantity which can be most
directly compared with the
experimental data \cite{bec_h}. 

\begin{figure}
\centerline{\epsfig{file=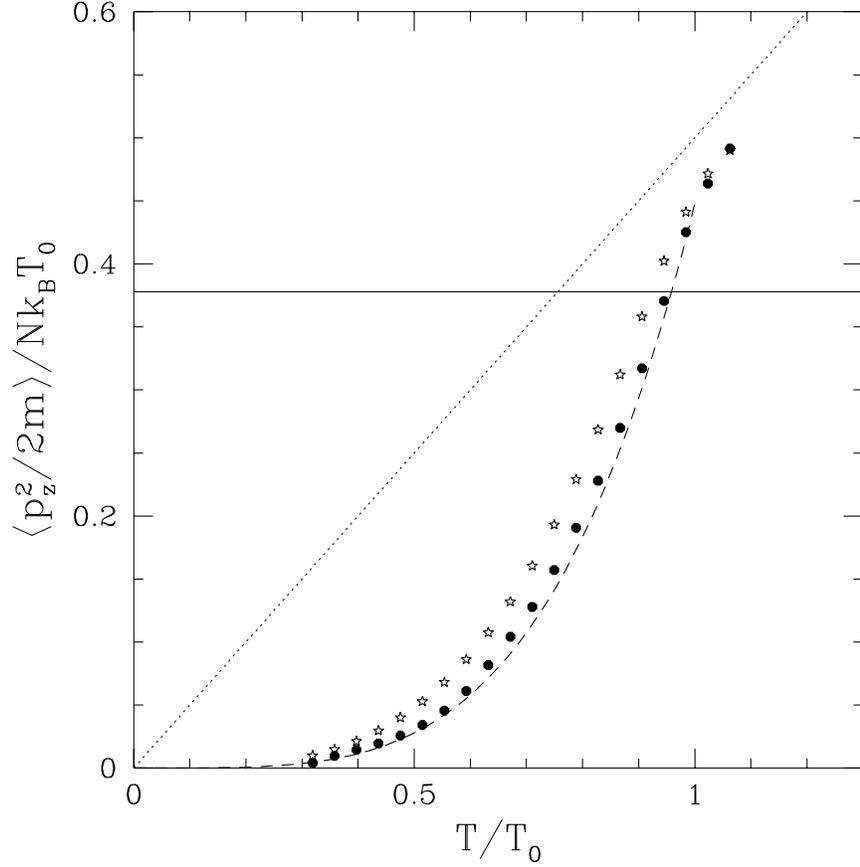,width=12cm}}
\caption{Kinetic energy versus temperature (in units of the
ideal-gas critical temperature $T_0$) for an
interacting Bose gas with
$N=10.5\cdot10^9$ at a scattering length $a=20a_H$
(stars) and at the hydrogen scattering length (filled circles).
These results are compared with the kinetic energy of the non-interacting
Bose gas (dashed curve) and of the classical gas (dotted curve). The
horizontal full line locates the kinetic energy measured in the hydrogen 
experiment~\protect\cite{bec_h}. }
\label{fig3}
\end{figure}

In Figure \ref{fig3} we show the behaviour of the kinetic energy of the
gas for two values of the scattering length, corresponding to the
experimental value $a=a_H$ for the hydrogen gas and to the stronger
interaction $a=20a_H$, as compared with the non-interacting quantum
and classical results. For the parameters corresponding to the
hydrogen experiment the effect of the interactions on the kinetic
energy is clearly quite negligible. Indeed, the dilution parameter for this
system is very small, $N^{1/6} a_H/a_{ho}\simeq0.001$ as already noted. 
On increasing the interaction strength we find that the kinetic energy
increases. This effect can be understood as coming from enhanced depletion of
the condensate.
There is instead a major effect of quantal statistics on the average
kinetic energy below the condensation temperature, as is seen
from Figure \ref{fig3}.

\begin{figure}
\centerline{\epsfig{file=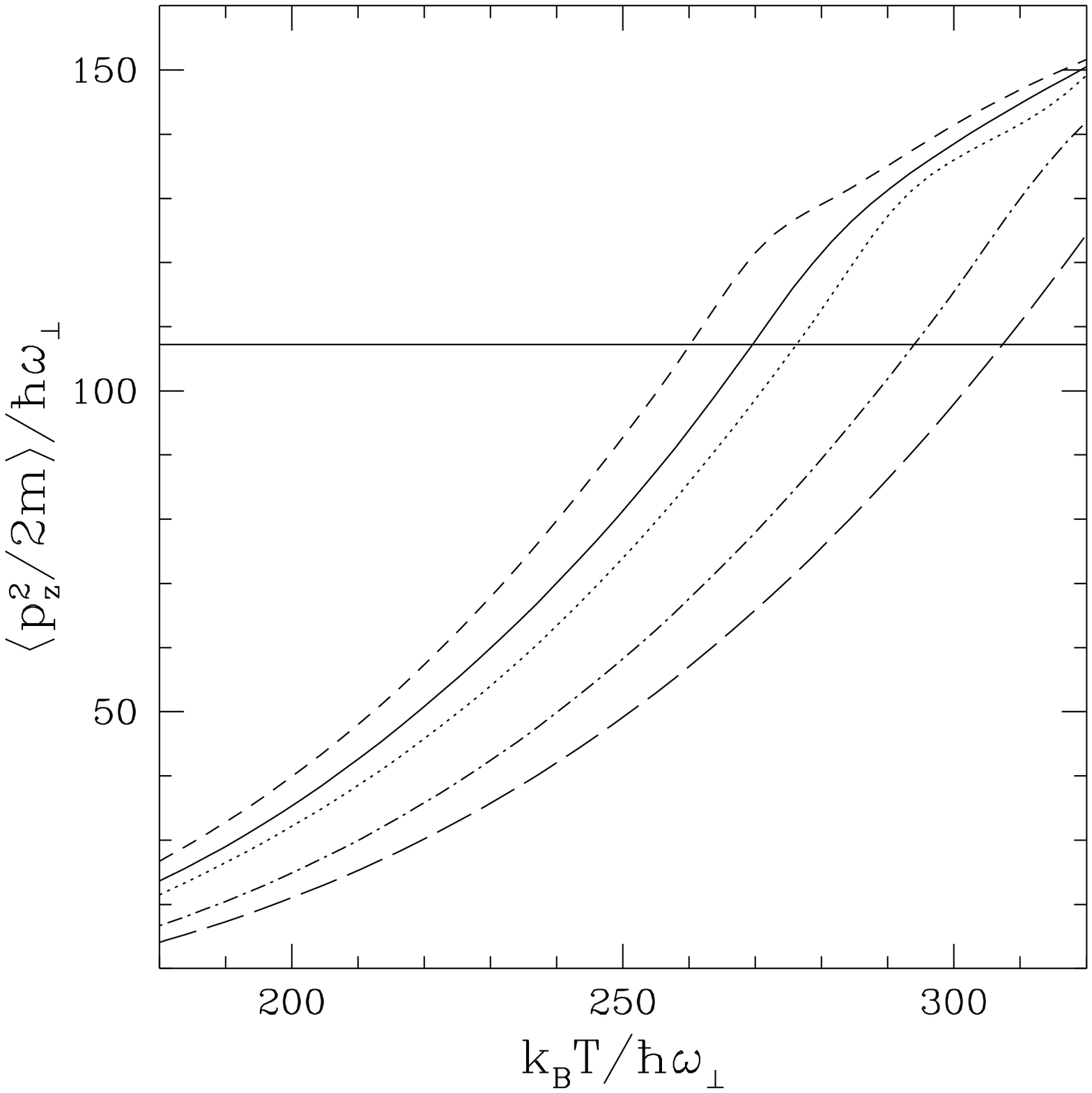,width=12cm}}
\caption{Kinetic energy versus temperature
for an interacting Bose gas at different values of the total number
of particles: $N=18.3\cdot 10^9$ (long-dashed), 
$N=15\cdot 10^9$ (dashed-dotted),
$N=11.6\cdot 10^9$ (dotted),
$N=10.5\cdot 10^9$ (continuous) and 
$N=9.5\cdot 10^9$ (short-dashed).
The horizontal full line locates the kinetic energy measured in the hydrogen
experiment~\protect\cite{bec_h}.}
\label{fig4}
\end{figure}

It is important to stress that in the interacting gas the behaviour of
the kinetic energy depends not only on the scattering length
but also on the total number of particles in the system.
This is illustrated in Figure \ref{fig4}. 
We have exploited this fact to test how far the semiclassical
model can quantitatively explain the experimental results 
of Fried {\it et al.}~\cite{bec_h} on the hydrogen gas. 
After choosing a value for the total number of particles
in the cloud within the experimental error bars,
we have fitted to experiment the calculated kinetic energy and
extracted from the model the
temperature of the cloud and  the condensate fraction. For a total
number of atoms equal to $N=10.5\cdot10^9$ we obtain $N_c/N=10\%$,
that is $N_c=1.05\cdot 10^9$ at
a temperature $T=50.3$ $\mu$K. The experimental values are
$N_c/N=6^{+6}_{-3}\%$ and $N_c=1.1\pm 0.6\cdot10^9$. 
Therefore, we conclude that the semiclassical model is compatible
with the experimental data for the momentum distribution of the thermal
cloud.

Our calculations also show that from kinetic energy data it is possible, 
through the semiclassical model to obtain a good estimate for
the temperature of the Bose cloud below the
critical temperature. In the intermediate- to low-temperature region
this procedure yields a result which differs significantly from the classical
result $E_{kin}^z=k_BT/2$. For the specific case of the hydrogen
experiment we estimate that the value of the
temperature corresponding to the measured kinetic energy is $T=50.3\,\mu$K,
that is about
$20\%$ higher than the value extrapolated from the classical result
(see again Figure~\ref{fig3}).

\section{Summary and conclusions}
In summary, the Doppler-sensitive part of the 1$S$-2$S$ two-photon 
absorption spectrum of Bose-condensed atomic hydrogen contains
quantitative information on thermodynamic properties, which can be
extracted from the available data by means of an analysis combining
a Thomas-Fermi treatment of the condensate with a semiclassical 
Hartree-Fock treatment of the thermal cloud. The spectral broadening
due to fluctuations in the density is negligible
and the spectral width is directly related to the kinetic energy
of the gas.

Within such a semiclassical two-fluid model we have studied the momentum
distribution and the mean kinetic energy of a Bose-condensed gas with
repulsive interactions. From the behaviour of the kinetic energy as a
function of temperature and of the coupling strength we have reached two 
main conclusions: ($i$) the interactions enhance the rate of depletion
of the condensate with increasing temperature, causing an increase of the 
kinetic energy; and ($ii$) the relation between kinetic energy and 
temperature is strongly non-linear, owing to the role of quantal
statistics as opposed to classical statistics.
It appears that the present model may yield useful improvements
in the analysis of experimental data.  

\acknowledgements
We thank Dr. C. Minniti for useful discussions and for her help
in the early stages of this work.

\newpage

\begin{thebibliography}{10}

\bibitem{ketterle_S_1}
J. Stenger, S. Inouye, A.~P. Chikkatur, D.~M. Stamper-Kurn, D.~E. Pritchard,
  and W. Ketterle, Phys. Rev. Lett. {\bf 82},  4569  (1999).

\bibitem{bec_h}
D.~G. Fried, T.~C. Killian, L. Willmann, D. Landhuis, S.~C. Moss, D. Kleppner,
  and T.~J. Greytak, Phys. Rev. Lett. {\bf 81},  3811  (1998).

\bibitem{baym}
G. Baym and C.~J. Pethick, Phys. Rev. Lett. {\bf 76},  6  (1996).

\bibitem{yvan_exp}
Y. Castin and R. Dum, Phys. Rev. Lett. {\bf 77},  5315  (1996).

\bibitem{cold_collision_shift}
T.~C. Killian, D.~G. Fried, L. Willmann, D. Landhuis, S.~C. Moss, T.~J.
  Greytak, and D. Kleppner, Phys. Rev. Lett. {\bf 81},  3807  (1998).

\bibitem{killian}
T.~C. Killian, Phys. Rev. A {\bf 61},  033611  (2000).

\bibitem{conti97}
A. Minguzzi, S. Conti, and M.~P. Tosi, J. Phys.: Condens. Matter {\bf 9},  L33
  (1997).

\bibitem{cote}
R. Cot\'e and V. Kharchenko, Phys. Rev. Lett. {\bf 83},  2100  (1999).

\bibitem{giorgini_scaling}
S. Giorgini, L.~P. Pitaevskii, and S. Stringari, Phys. Rev. Lett. {\bf 78},
  3987  (1997).

\bibitem{muntsa}
F. Dalfovo, S. Giorgini, M. Guilleumas, L.~P. Pitaevskii, and S. Stringari,
  Phys. Rev. A {\bf 56},  3840  (1997).

\bibitem{markus_hf}
M. Holzmann, W. Krauth, and M. Naraschewski, Phys. Rev. A {\bf 59},  2956
  (1999).

\bibitem{cornell_thermod}
J.~R. Ensher, D.~S. Jin, M.~R. Matthews, C.~E. Wieman, and E.~A. Cornell, Phys.
  Rev. Lett. {\bf 77},  4984  (1996).

\bibitem{jamieson}
M.~J. Jamieson, A. Dalgarno, and M. Kimura, Phys. Rev. A {\bf 51}, 2626
(1995).

\bibitem{chou2}
T. Chou, C.~N. Yang, and L. You, Phys. Rev. A {\bf 55},  1179  (1997).

\end{thebibliography}
\end{document}